\documentclass[aip,amsmath,amssymb]{revtex4-1}
\usepackage{amsmath}
\usepackage{color}
\usepackage{amssymb}
\usepackage[caption=false]{subfig}
\usepackage{graphicx}% Include figure files
\usepackage{dcolumn}% Align table columns on decimal point
\usepackage[utf8]{inputenc}
\usepackage{float}
\definecolor{orange}{rgb}{1,0.5,0}

\begin{document}

\title{\textcolor{black}{Frozen-Density Embedding Theory based simulations using experimental electron densities for the environment}}

% Authors' names and addresses. Use \cauthor for the main (contact) author.
     % Use \author for all other authors. Use \aff for authors' affiliations.
     % Use lower-case letters in square brackets to link authors to their
     % affiliations; if there is only one affiliation address, remove the [a].

\author{Niccol\`{o} Ricardi}%
\email{Niccolo.Ricardi@unige.ch}
\affiliation{%
University of Geneva,
Department of Physical Chemistry,
30, Quai Ernest-Ansermet,
CH-1211 Gen\`{e}ve 4,
Switzerland
}
\author{Michelle Ernst}%
\email{Michelle.Ernst@dcb.unibe.ch}
\affiliation{%
University of Bern, 
Freiestr. 3, 3012 Bern
Switzerland
}
\author{Piero Macchi}%
\email{Piero.macchi@polimi.it}
\affiliation{%
Polytechnic of Milan, 
Department of Chemistry, Materials and Chemical Engineering,
via Mancinelli 7, Milano 20131, Italy
}
\author{Tomasz A. Wesolowski}%
% \email{Tomasz.Wesolowski@unige.ch, Niccolo.Ricardi@unige.ch}
\email{Tomasz.Wesolowski@unige.ch}
\affiliation{%
University of Geneva,
Department of Physical Chemistry,
30, Quai Ernest-Ansermet,
CH-1211 Gen\`{e}ve 4,
Switzerland
}%

\date{\today}% It is always \today, today,

\begin{abstract}
The basic idea of Frozen-Density Embedding Theory (FDET) is the constrained minimisation of the Hohenberg-Kohn density functional $E^{HK}[\rho]$ performed using the auxiliary functional $E_{v_{AB}}^{FDET}[\Psi_A,\rho_B]$, where $\Psi_A$ is the embedded $N_A$-electron wave-function and $\rho_B(\vec{\mathrm{r}})$ a non-negative function in real space integrating to a given number of electrons $N_B$. 
This choice of independent variables in the total energy functional $E_{v_{AB}}^{FDET}[\Psi_A,\rho_B]$ makes it possible to treat  the corresponding two components of the total density using different methods in multi-level simulations. 
We demonstrate, for the first time, the applications of FDET using $\rho_B(\vec{\mathrm{r}})$ reconstructed from  X-ray diffraction data on a molecular crystal.
 For eight hydrogen-bonded clusters involving a chromophore (represented with $\Psi_A$)  and the glycylglycine molecule (represented as 
 $\rho_B(\vec{\mathrm{r}})$), FDET is used to derive excitation energies.
 It is shown that experimental densities are suitable to be used as $\rho_B(\vec{\mathrm{r}})$  in FDET based simulations.
\end{abstract}

\maketitle

\section{Introduction}
Frozen-Density Embedding Theory (FDET) is the Hohenberg-Kohn theorems based formal framework for multi-level simulations.~\cite{Wesolowski2004}
The total electron density is built up from two components $\rho_A(\vec{\mathrm{r}})$ and $\rho_B(\vec{\mathrm{r}})$ of which only the first is constructed from quantum-mechanical descriptors. FDET was originally formulated for  variational methods used to obtain such descriptors of the embedded species as: a)  non-interacting reference system described with a Kohn-Sham determinant~\cite{Wesolowski1993}, b) interacting system described with a multi-determinant wave-function~\cite{Wesolowski2008}, and c) one \textcolor{black}{particle density}-matrix~\cite{Pernal2009}. Extension of FDET for non-variational methods has been recently formulated~\cite{Zech2019}. Extensions of FDET for excited states can be made based either on response theory  for non-interacting~\cite{Wesolowski2004} or interacting~\cite{Hofener2012b} systems.  Another possibility to describe excited states relies on the Perdew-Levy theorem on extrema of the ground-state energy functional~\cite{Perdew1985}. It makes it possible to interpret other-than-the-lowest energy stationary embedded wave-functions obtained in FDET as excited states as pointed out by Khait and Hoffmann~\cite{Khait2010}. 
In any of these variants of FDET, the embedded wave-function depends on the chosen $\rho_B(\vec{\mathrm{r}})$. Several computational methods sharing with FDET some elements but differing in some key aspects such as the choice of independent variables,  self-consistency between embedding potential and the embedded wave-function,  locality of the embedding potential, etc. have been developed in various groups. We address the reader to reviews concerning - besides the methods based on FDET - also related computational approaches~\cite{Wang2000,Wesolowski2006,Jacob2014,Wesolowski2015,Krishtal2015}. 

At the present state of development of approximations for the FDET embedding functional (see Eq.~\ref{eq:v_emb}),
applications of FDET are limited to such systems where  $\rho_A(\vec{\mathrm{r}})$ and $\rho_B(\vec{\mathrm{r}})$ do not overlap significantly~\cite{Wesolowski1996c,Bernard2008}. As a rule of thumb, FDET based methods are only applicable to such cases where the environment is not covalently bound to the embedded species~\cite{Gotz2009,Goodpaster2010,Fux2010a}.  
In such cases, the overlap between $\rho_A(\vec{\mathrm{r}})$ and $\rho_B(\vec{\mathrm{r}})$ is small and simple local- and semi-local approximations are sufficiently accurate.
FDET based simulations can be seen as the variant of QM/MM simulations, in which the modeller decides about the procedure to generate  
$\rho_B(\vec{\mathrm{r}})$ instead of parametrising the force-field parameters describing the energy contributions due to the interactions between the quantum system and its environment.
Various system- and property specific protocols for generating $\rho_B(\vec{\mathrm{r}})$ for FDET based simulations are possible.
Some examples of different treatments of the environment density are given below.
If the environment comprises several weakly bound molecules, the corresponding $\rho_B(\vec{\mathrm{r}})$ can be obtained either  from quantum mechanical calculations for the whole cluster comprising all molecules in the environment or,  in a simplified manner, 
as a superposition of molecular densities derived from some quantum-mechanical method~\cite{Wesolowski1994,Humbert-Droz2014}. If $\rho_B(\vec{\mathrm{r}})$ is localised in a pre-defined part of the space, the effect of electronic polarisation of the environment by the embedded species can be taken into account by optimising also $\rho_B(\vec{\mathrm{r}})$~\cite{Wesolowski1996a} or by "pre-polarising" it using simpler techniques ~\cite{Zbiri2004,Ricardi2018}. 
FDET can also be used for setting up a multi-physics simulation in which  $\rho_B(\vec{\mathrm{r}})$ represents a statistical ensemble averaged electron density ($<\rho_B>(\vec{\mathrm{r}})$) represented as a continuum derived using classical statistical-mechanics based approaches~\cite{Kaminski2010,Laktionov2016}.  Such methods are especially useful for studying electronic structure of solvated molecules~\cite{Shedge2014b}.

The above examples show clearly that the choice of the procedure to generate $\rho_B(\vec{\mathrm{r}})$ is the key element of any FDET based simulation.
This can be made in an "automatic" way by making some system-independent procedures/choices/approximations or made in a system dependent manner involving user provided information about $\rho_B(\vec{\mathrm{r}})$ such as:  a) using as $\rho_B(\vec{\mathrm{r}})$ the ground-state density of some system obtained without putting any information about embedded species, b) localising 
$\rho_B(\vec{\mathrm{r}})$ in a pre-defined region of space by choosing a limited set of atom-centred basis functions, c) allowing it to spread over the whole system, d) optimising $\rho_B(\vec{\mathrm{r}})$ by means of the "freeze-and-thaw" minimisation of the total energy~\cite{Wesolowski1996a} e) or any combination of the above. 
In principle, the density $\rho_B(\vec{\mathrm{r}})$ obtained from the unique partitioning of the total density using the approach developed by Carter and collaborators~\cite{Huang2011a,Huang2011b} could be used as a  possible "automatic" procedure to generate $\rho_B(\vec{\mathrm{r}})$ in FDET.

The strategy, in which $\rho_B(\vec{\mathrm{r}})$ is obtained from quantum mechanical calculations for the environment only, i.e., 
in the absence of the embedded species, is particularly attractive. \textcolor{black}{
Both our experience and work by other researchers show that obtaining $\rho_B(\vec{\mathrm{r}})$ from an isolated calculation yields the dominant contribution to the complexation induced shifts of the excitation energies especially for excitations with shifts of large magnitude (see~\cite{Fradelos2011b,Zech2018,Ricardi2018}).
Daday et al. showed that the effects of the optimisation of $\rho_B(\vec{\mathrm{r}})$, either for the ground- or both ground- and excited-state, are secondary, albeit numerically non-negligible: the excitation energy for methylenecyclopropene solvated by 17 water molecules (which has a reference shift of 0.86~eV obtained from  CASPT2 calculations for the whole cluster) is fairly reproduced (0.82~eV) with such a choice for $\rho_B(\vec{\mathrm{r}})$~\cite{Daday2014}. In the case of  $n-\pi^*$ excitations for acrolein in water, the corresponding shifts  are 1.42~eV and 1.10~eV. Although the  difference between the FDET with such choice of $\rho_B(\vec{\mathrm{r}})$ and reference shifs  cannot be  attributed to the ``neglect of the electronic polarisation of the environment'' within the formal framework of FDET, the optimisation of $\rho_B(\vec{\mathrm{r}})$~\cite{Daday2014} or pre-polarisation~\cite{Ricardi2018} usually reduce this difference.
This secondary importance of the explicit treatment of the polarisation of the environment is due to the variational character of FDET and the fact that the partitioning of the total density of the complex into $\rho_A(\vec{\mathrm{r}})$ and $\rho_B(\vec{\mathrm{r}})$ is not unique in exact FDET, resulting in a better capacity to approach the exact total density (see the discussion in~\cite{Wesolowski2015,Humbert-Droz2014}).}  

The present work concerns yet another possibility to generate $\rho_B(\vec{\mathrm{r}})$ for FDET simulations of embedded species in a given environment consisting of non-covalently bound molecules, in which $\rho_B(\vec{\mathrm{r}})$ is obtained from experimental data concerning a different system: a molecular crystal of the environment molecule.
Recent years brought a number of works showing that both electron densities~\cite{Hansen1978} and wavefunctions~\cite{Jayatilaka2012} can be reconstructed from X-ray diffraction data. 
It is tempting, therefore, to explore these new possibilities generate $\rho_B(\vec{\mathrm{r}})$ for the use in FDET based simulations.  We have to underline that several approximations may undermine the use of X-ray based densities for FDET. 
The most important issues can be listed as it follows: a) the link to any experimental quantity is of course affected by eperimental errors, which are unavoidable and may affect both precision and accuracy; b) the electron density and wavefunction that are extracted from experiment are static, whereas atoms are not steady in the crystal; c) the  sampling of the diffraction in the reciprocal space is necessarily incomplete; d) only the intensity of the diffracted ray is measured, but not the phase; e) the crystal sample is imperfect.
For these reasons, the possibility to use experimental densities as $\rho_B(\vec{\mathrm{r}})$  in FDET, hinges critically on robust and numerically stable protocols to generate such densities and it is hence important to investigate the dependence of FDET results on such procedures.
The state of the art in the wavefunction and electron density recustruction from X-ray diffraction data is encouraging this attempt. In particular the above mentioned pitfalls may be tackled as follows: a) modern instrumentation enables measuring the diffraction intensities with high precision; b) the deconvolution of thermal motion is reliable, if the measurements are carried out at sufficiently low temperature and if the resolution of the diffraction is sufficiently large; c) complementary input from theory can compensate for the missing information; d) appropriate modelling enables the phasing of the diffracted rays; e) data correction from ideal kinematic theory of diffraction allows for sufficiently accurate data.
The present work reports an exploratory study on the use of densities from X-ray restricted wavefuntions in FDET.

Concerning a particular variant of FDET and system to be investigated, we have chosen to evaluate the excitation energies obtained from LinearizedFDET~\cite{Wesolowski2014,Zech2015} for several organic chromophores, each, hydrogen bonded to its environment. 
Our extensive benchmarking of the performance of FDET for such cases indicate that the errors of FDET excitation energies due to the used approximations for the explicit density functional for non-electrostatic components of the FDET embedding potential (see the next section) are small.
In a benchmark set of embedded organic chromophores, the average deviation from the reference amounts to about 0.04~eV~\cite{Zech2018}. 
This magnitude of the deviation defines the threshold for complexation-induced shifts in the excitation energy above which analysis of the dependence on the shift on $\rho_B(\vec{\mathrm{r}})$ is meaningful. In the embedded chromophores chosen for the present study these shifts vary 
between 0.15 to 0.6 eV.

\section{Embedded chromophores}
Concerning the molecules for which  $\rho_B(\vec{\mathrm{r}})$ is generated, we have chosen glycylglycine (GlyGly). For this exploratory study, it is crucial that the molecule(s) corresponding to  $\rho_B(\vec{\mathrm{r}})$ are capable to form hydrogen bonds with the chromophore.    GlyGly satisfies this condition. Moreover, the molecular density of GlyGly reconstructed from X-ray diffraction data reflects the features arising from intermolecular hydrogen bonds present in the crystal~\cite{Genoni2018}.

\begin{figure}
\centering
\includegraphics[width=0.8\textwidth]{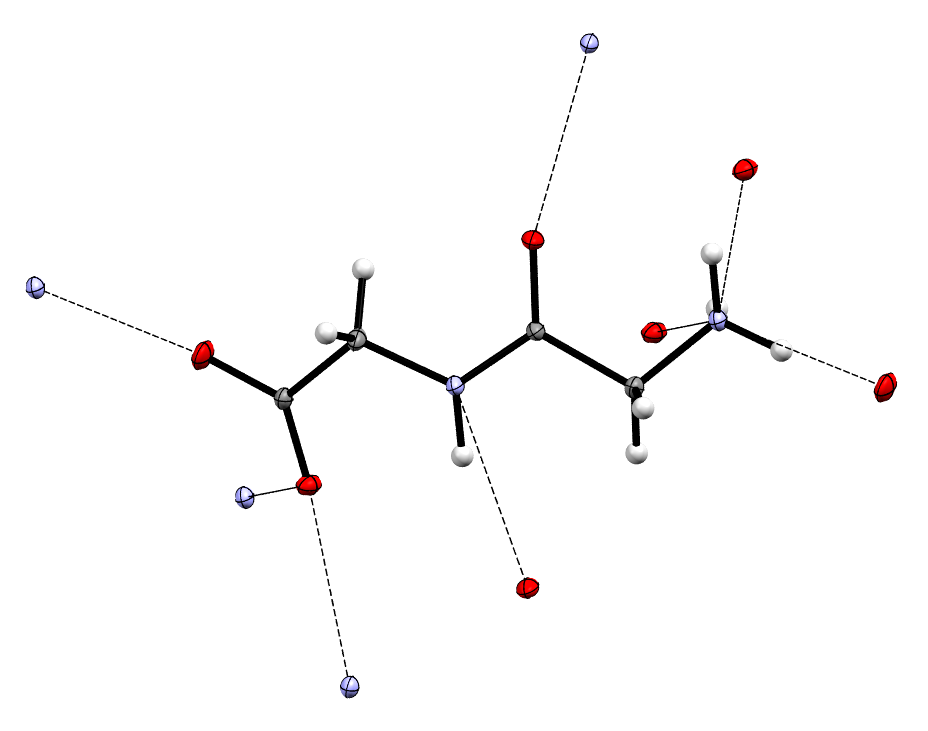}
\caption{The hydrogen-bonding pattern for the glycylglycine molecule in the crystal taken from Ref.~\cite{DosSantos2014}. Only nearest atoms involved in hydrogen bonding are shown: oxygen (red) and nitrogen (blue).} 
\label{Figure_GlyGly}
\end{figure}

 Figure~\ref{Figure_GlyGly} shows the GlyGly molecule together with its nearest neighbours in the crystal.

\begin{figure}
\centering
\includegraphics[width=0.9\textwidth]{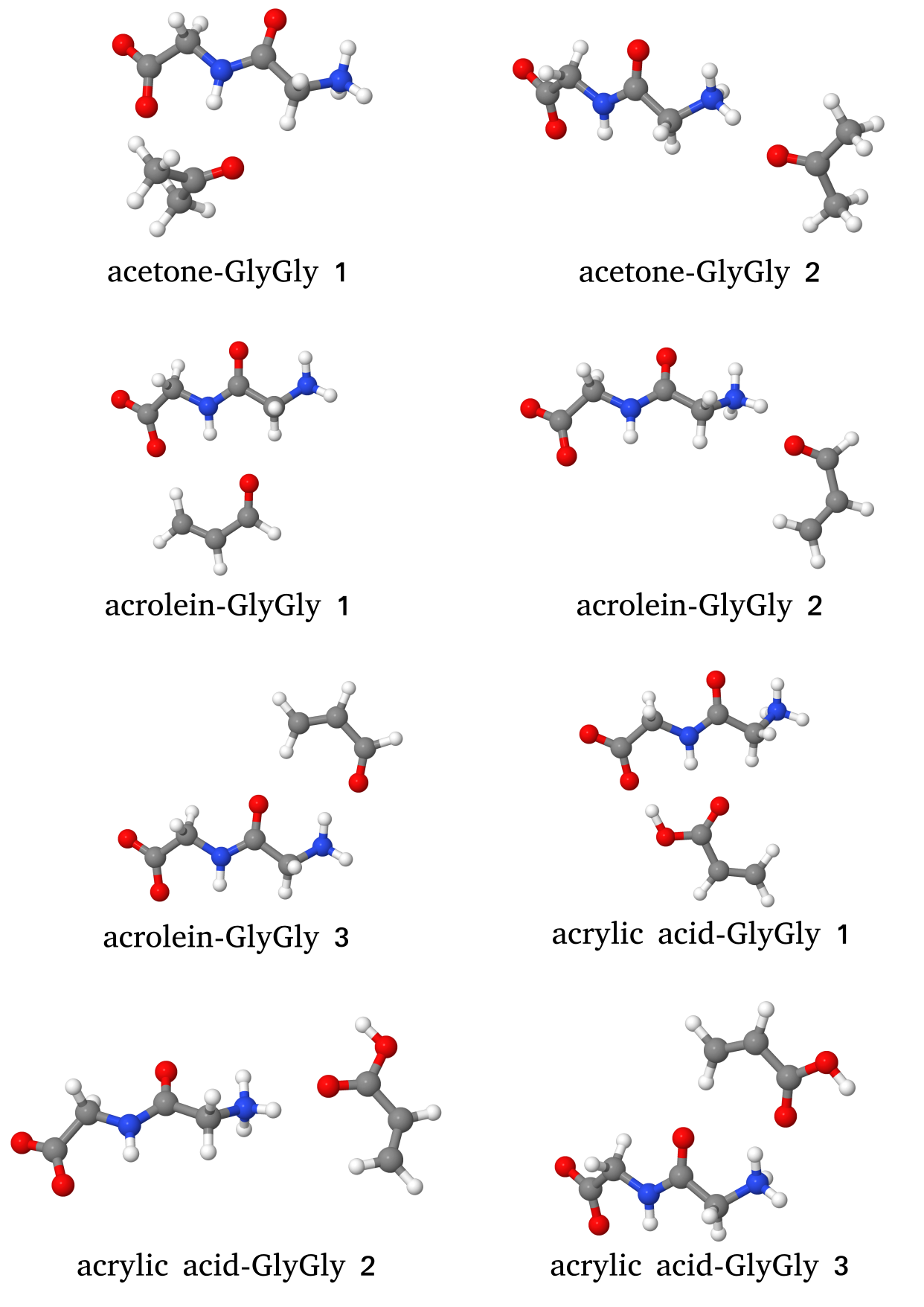}
\caption{The complexes of three different chromophores with glycylglycine.}
\label{Figure_complexes}
\end{figure}

The densities reconstructed from experimental data on glycylglycine, are  used in the present work as $\rho_B(\vec{\mathrm{r}})$ in FDET calculations 
of excitation energies for eight different hydrogen-bonded complexes formed by one organic chromophore (acrolein, acrylic acid, or acetone) and glycylglycine. Figure~\ref{Figure_complexes} shows the considered clusters. 

The hydrogen-bonding networks shown in Figures~\ref{Figure_GlyGly} and~\ref{Figure_complexes} are not the same. In the crystal, all donors and acceptors  are involved in hydrogen bonding which is not the case in the investigated clusters. Nevertheless, each individual hydrogen bond in the clusters has its corresponding partner in the crystal. It can be expected, therefore, that the effect of the hydrogen bonding on $\rho_B(\vec{\mathrm{r}})$ in the cluster is also reflected in the density obtained from the crystal.

\section{FDET approach to multi-level simulations}
For a system comprising N$_{AB}$ electrons in an external potential  $v_{AB}(\vec{\mathrm{r}}\,)$, the functional $E_{v_{AB}}^{FDET}\left[\Psi_{A}^{},\rho_{B}^{}\right]$ is defined to satisfy by construction the following relation:
\begin{equation}\label{eq:FDET_equality}
 \min_{\Psi_A} {E_{v_{AB}}^{FDET}}\left[\Psi_{A},\rho_{B}^{}\right] = {E_{v_{AB}}^{FDET}}\left[\Psi_{A}^{o},\rho_{B}^{}\right]=
 E_{v_{AB}}^{HK}[\rho_A^{o}+\rho_B],
\end{equation}
where $E_{v_{AB}}^{HK}[\rho]$ is the Hohenberg-Kohn ground-state energy functional~\cite{Hohenberg1964} and
$
\displaystyle
\rho_{A}^{o}({\vec{\mathrm{r}}\,}) = 
\big\langle\Psi_{A}^{o}\big|
\sum_i^{N_A}\delta(\vec{\mathrm{r}}-\vec{\mathrm{r}}_i)
\big|\Psi_{A}^{o}\big\rangle
$.

By virtue of the second Hohenberg-Kohn theorem, Eq. \ref{eq:FDET_equality} leads to:
\begin{equation}\label{eq:FDET_inequality}
 {E_{v_{AB}}^{FDET}}\left[\Psi_{A}^{o},\rho_{B}^{}\right]\ge E_0
\end{equation}
where $E_0=E_{v_{AB}}^{HK}[\rho_0]$ and $\rho_{0}^{}(\vec{\mathrm{r}})$ is the ground-state energy and density of the total system. 
Equality is reached for a large class of densities $\rho_{B}({\vec{\mathrm{r}}\,})$: 
\begin{equation}\label{eq:FDET_equality1}
 {E_{v_{AB}}^{FDET}}\left[\Psi_{A}^{o},\rho_{B}^{}\right]
 = E_0\;\;\;\mathrm{if}\;\;\;
\forall \vec{\mathrm{r}} \left(\rho_{0}^{}(\vec{\mathrm{r}})>\rho_B^{}(\vec{\mathrm{r}})\right).
\end{equation}

Using conventional density functionals representing components of the total energy,
and arbitrary partitioning of the external potential $v_{AB}({\vec{\mathrm{r}}\,}) = v_{A}({\vec{\mathrm{r}}\,})+v_{B}({\vec{\mathrm{r}}\,})$, leads to the form of
${E_{v_{AB}}^{FDET}}\left[\Psi_{A}^{},\rho_{B}^{}\right]$ more suitable for further discussions:
\begin{eqnarray}
 {E_{v_{AB}}^{FDET}}\left[\Psi_{A}^{},\rho_{B}^{}\right] &=&
\left\langle\Psi_{A}^{}\left|\hat{H}_{A}^{}\right|\Psi_{A}^{}\right\rangle
+ V_{B}^{}\left[\rho_{A}^{}\right]
 + J_{AB}^{}\left[\rho_{A}^{},\rho_{B}^{}\right] \nonumber\\ 
&+& E_{xcT}^{nad}\left[\rho_{A}^{},\rho_{B}^{}\right]  +  \Delta F\left[\rho_{A}\right] \label{eq:FDET_energy}\\
 &+& E_{v_{B}}^{HK}\left[\rho_{B}^{}\right] +
V_{A}^{}\left[\rho_{B}^{}\right] + V_{N_AN_B}\nonumber
\end{eqnarray}
where
\begin{eqnarray}
\nonumber
V_{A}^{}\left[\rho_{B}^{}\right]&=&\int v_{A}({\vec{\mathrm{r}}\,}) \rho_{B}({\vec{\mathrm{r}}\,}) d{\vec{\mathrm{r}}\,}\\\nonumber
V_{B}^{}\left[\rho_{A}^{}\right]&=&\int v_{B}({\vec{\mathrm{r}}\,}) \rho_{A}({\vec{\mathrm{r}}\,}) d{\vec{\mathrm{r}}\,}\\\nonumber
J_{AB}^{}\left[\rho_{A}^{},\rho_{B}^{}\right]&=&\nonumber
\int \int 
\frac{\rho_{A}^{}({\vec{\mathrm{r}}\,})\rho_{B}^{}({\vec{\mathrm{r}}\,}')}{\left|\vec{\mathrm{r}}-{\vec{\mathrm{r}}\,}'\right|}\mathrm{d}{\vec{\mathrm{r}}\,}'
{d}{\vec{\mathrm{r}}\,}
\end{eqnarray}
and $V_{N_AN_B}$ is the interaction energy between the nuclei defining $v_{A}({\vec{\mathrm{r}}\,})$ and $v_{B}({\vec{\mathrm{r}}\,})$.
The non-additive bi-functional $E_{xcT}^{nad}\left[\rho_{A}^{},\rho_{B}^{}\right]$ is related to the functionals $E_{xc}[\rho]$ and $T_s[\rho]$ defined in the constrained search formulation of the Kohn-Sham formalism ~\cite{Levy1979}. It is defined as:
\begin{eqnarray}E_{xcT}^{nad}\left[\rho_{A}^{},\rho_{B}^{}\right]&=&E_{xc}\left[\rho_{A}^{}+\rho_{B}^{}\right]-E_{xc}\left[\rho_{A}^{}\right]-E_{xc}\left[\rho_{B}^{}\right]+\nonumber\\
&+& T_{s}\left[\rho_{A}^{}+\rho_{B}^{}\right]-T_{s}\left[\rho_{A}^{}\right]-T_{s}\left[\rho_{B}^{}\right]
\end{eqnarray}
The functional  $\Delta F\left[\rho\right]$ on the other hand 
 depends on the form of the wavefunction $\Psi$ used in Eq.\,\ref{eq:FDET_equality} and also 
 is defined via the constrained search ~\cite{Wesolowski2008}. 
For instance, if $\Psi_{A}^{}$  is a single determinant ($\Phi$),  
 it reads:
\begin{eqnarray}
\Delta F^{SD}[\rho]&=&
%\min_{\Psi\longrightarrow \rho}\left<\Psi\vert \hat{T}_{N}+\hat{V}^{ee}_{N}\vert \Psi\right>&=&\left<\Psi^{opt}[\rho]\vert \hat{T}_{N_A}+\hat{V}^{ee}_{N_A} \vert \Psi^{opt}[\rho]\right>=T[\rho]+V_{ee}[\rho]\nonumber\\
\min_{\Phi\longrightarrow \rho}\left<\Phi\vert {\hat{T}_{N_A}+\hat{V}^{ee}_{N_A}}\vert \Phi\right>- T[\rho]-V_{ee}[\rho] \\
&=& \left<\Phi^{o}[\rho]\vert {\hat{T}_{N_A}+\hat{V}^{ee}_{N_A}} \vert \Phi^{o}[\rho]\right>-T[\rho]-V_{ee}[\rho]\nonumber 
= E_c[\rho] \nonumber
\end{eqnarray}
and is just the correlation functional defined in constrained-search formulation of density functional theory ~\cite{Levy1979,Baroni1983}. 
For  $\Psi$ of the full CI form, $\Delta F^{FCI}\left[\rho\right]=0$ by definition. 

Euler-Lagrange optimisation of $\Psi_{A}^{}$ leads to the Schr\"odinger-like equation: 
\begin{eqnarray}
\left(\hat{H}_{A}^{}+\hat{\upsilon}_{emb}^{}\right)\Psi_{A}^{}=\lambda\Psi_{A}^{}
\label{ps_schr}
\end{eqnarray}
where
\begin{eqnarray}\label{eq:v_emb}
 v_{emb}^{}[\rho_{A}^{},\rho_{B}^{},v_{B}^{}](\vec{\mathrm{r}})
 &=& v_{B}^{}(\vec{\mathrm{r}}) + \int\frac{\rho_{B}^{}({\vec{\mathrm{r}}\,}')}{\left|\vec{\mathrm{r}}-{\vec{\mathrm{r}}\,}'\right|}\mathrm{d}{\vec{\mathrm{r}}\,}'\\
&+& v_{xcT}^{nad}[\rho_{A},\rho_{B}](\vec{\mathrm{r}})
+ v_{F}[\rho_{A}](\vec{\mathrm{r}})\nonumber
\end{eqnarray}
with $v_{xcT}^{nad}[\rho_{A},\rho_{B}](\vec{\mathrm{r}})$, 
and   $v_{F}[\rho_A](\vec{\mathrm{r}})$ 
being  the first functional derivatives of $E_{xcT}^{nad}\left[\rho^{},\rho_{B}^{}\right]$ and $\Delta F\left[\rho\right]$, respectively.

The lowest energy solution of Eq.\,\ref{ps_schr} will be denoted as $\Psi_{A}^{EL}$. 
Note that the energy is given not by the Lagrange multiplier $\lambda$ but in Eq.\,\ref{eq:FDET_energy}.
For exact density functionals, any variational method can be used to obtain $\Psi_{A}^{EL}$ and the corresponding density $\rho_A^{EL}(\vec{\mathrm{r}})$, which satisfy be construction the basic FDET equality given in Eq.~\ref{eq:FDET_equality}.

\subsection{Reconstruction of $\rho_B(\vec{\mathrm{r}})$ from X-ray diffraction data}

X-ray restrained wavefunctions (XRW), in literature commonly (but incorrectly) termed X-ray constrained wavefunctions, 
were initially developed by Jayatilaka~\cite{Jayatilaka2012,Jayatilaka2001,Grimwood2001}.
In XRW, instead of applying the variational principle, like in conventional SCF, a special functional $L$ is defined, based on a classical Hamiltonian and a function of the square difference between calculated and experimentally measured structure factors, which ideally distributes with a $\chi^2$ statistics. \textcolor{black}{Where:
\begin{equation}
 \chi^2(F^{exp}) = \frac{1}{N_r - N_p} \sum\limits_k^{N_r} \frac{\left((F_k-F^{exp}_k\right)^2}{\sigma^2_k}
 \end{equation}
 with $N_r$ and $N_p$ being respectively the number of experimental data and parameters in the model, $(F_k-F^{exp}_k)$ is the difference between the structure factors from the trial wavefunction and the experimental ones, and $\sigma_k$ is the experimental standard deviation. Thus, the minimization of $L$ implies finding the minimal energy AND the best agreement with experiment. Of course, this cannot be simoultaneously achieved and a parameter $\lambda_j$ must be defined in order to weight the two parts of the functional. Therefore, the functional takes the form:
\begin{equation}
 L=E+\lambda_J \chi^2
\end{equation}
}
This procedure allows to construct molecular wavefunctions from experimental observations in crystals. By increasing $\lambda_j$, both long and short range interactions in the crystal are progressively taken into account.
In this work, we used structure factors measured for GlyGly to calculate X-ray restrained wave-functions with $\lambda_J$ values from 0.0 to 1.0, as for higher values the SCF procedure does not converge. We stress that such value of $\lambda_J$=1.0 has no specific meaning because the electronic energy of the Hamiltonian and the electron density difference in the $\chi^2$ function have two different units, thus $\lambda_J$ is not dimensionsless but it depends on the number of electrons, the molecular volume, and the diffraction resolution. Moreover, the structure factors in the $\chi^2$ function are weighted by the variance of their measurement statistics.
The aforementioned wave-functions were then used to calculate $\rho_B(\vec{\mathrm{r}})$.

\subsection{Computational Details}
The following approximations were used in the reported FDET calculations:
\textit{i}) ADC(2) treatment~\cite{Schirmer1982} of correlation for embedded $N_A$ electrons as implemented in  Ref.~\cite{Prager2016},  
\textit{ii}) decomposable approximations for $v_t^{nad}[\rho_A,\rho_B](\vec{\mathrm{r}})$ (LDA) and $v_{xc}^{nad}[\rho_A,\rho_B](\vec{\mathrm{r}})$ 
(note that in the LinearizedFDET used here approximations for the energy components $E_{xcT}^{nad}[\rho_A,\rho_B]$ and $\Delta F[\rho_A]$ are not used at all), 
\textit{iii}) neglect the $v_F[\rho_A]$ contribution to the embedding potential, 
\textit{iv}) monomer expansion of  $\rho_A(\vec{\mathrm{r}})$ 
(only atomic basis sets centred on the chromophore), 
\textit{v})  monomer expansion of  $\rho_B(\vec{\mathrm{r}})$ 
(only atomic basis sets centred on GlyGly),
\textit{vi}) chromophore-independent generation of  $\rho_B(\vec{\mathrm{r}})$ using one of the following methods for the isolated GlyGly:  Hartree-Fock, first-order M\o ller-Plesset perturbation theory, Kohn-Sham with \textcolor{black}{PBE~\cite{Perdew1996}} approximation for $E_{xc}[\rho]$, CCSD~\cite{Bartlett1981}, or the reconstruction from experimental structure factors (see the next section). 

The FDET results for each cluster are compared to the reference obtained from ADC(2) calculations. 
The reported reference shifts in 
the excitation energy are evaluated as $\Delta\epsilon_{ref}=\epsilon_{AB}^{ADC(2)}-\epsilon_{A(B)}^{ADC(2)}$, 
where $AB$ denotes the complex and $A(B)$ denotes the chromophore alone but with the basis set  expanded by the functions localised on GlyGly (similarly as it is made in the counterpoise technique of Boys and Bernardi~\cite{Boys1970} for intermolecular interaction energy). 
In all calculations including also the reconstruction of the electron density of GlyGly from the X-ray structure factors,  the 6-311G basis set was used.

At ${\lambda_J=0}$, the wave-function obtained from X-ray structure data is just the Hartree-Fock molecular wave-function.
The numerical results should be identical regardless the used software to generate $\rho_B(\vec{\mathrm{r}})$.
We used this fact to check the numerical soundness of the procedures to export-import densities $\rho_B(\vec{\mathrm{r}})$ obtained with different software. Tonto~\cite{Grimwood2003} was used for X-ray restrained wavefunction calculations, Psi-4 ~\cite{Parrish2017} for to generate the CCSD GlyGly density,and Q-Chem~\cite{Qchem4}\textcolor{black}{, with its ADCMAN~\cite{Wormit2014} and FDEMAN~\cite{Prager2016} modules,} for all other calculations, including FDET/ADC(2) ones. 

Throughout the text, $\epsilon_{emb}[\rho_B^{method}]$ (and $\Delta\epsilon_{emb}[\rho_B^{method}]$)  denote the FDET derived excitation energy (and environment induced shift), where the subscript in $\rho_B^{method}(\vec{\mathrm{r}})$ specifies the method 
used to generate  $\rho_B(\vec{\mathrm{r}})$.

\section{Results}

\begin{figure}
\centering
\includegraphics[width=0.9\textwidth]{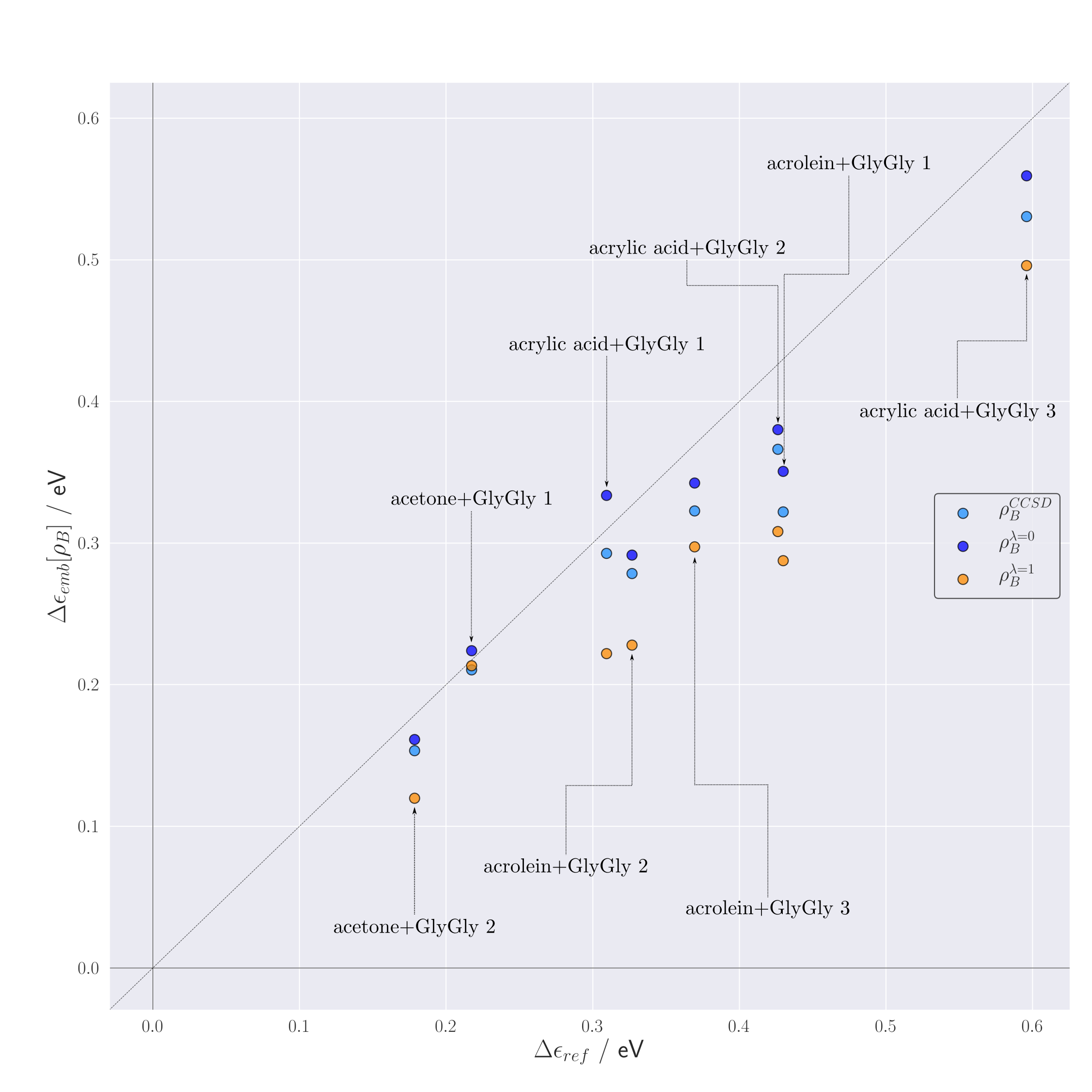}
\caption{Complexation induced shifts of the  excitation energy ($\Delta\epsilon_{emb}[\rho_B]$) for eight chromophores hydrogen bonded to GlyGly.
For each complex, FDET calculations (embedded ADC(2)) \textcolor{black}{using as $\rho_B(\vec{\mathrm{r}})$} the electron density of GlyGly obtained from three different methods: Hartree-Fock \textcolor{black}{($\lambda_J=0$)} or CCSD for the isolated GlyGly or density reconstructed from X-ray structure factors for GlyGly molecular crystal at $\lambda_J=1$.  
Reference values ($\Delta\epsilon^{ref}$) are obtained from ADC(2) calculations for the whole complex.}
\label{Figure_diagonal}
\end{figure}

For eight considered clusters, the lowest excitation energies obtained from FDET/ADC(2) calculations ($\epsilon_{emb}[\rho_B]$) using several choices for $\rho_B(\vec{\mathrm{r}})$ are shown in
Figure~\ref{Figure_diagonal} together with the corresponding reference supermolecular ADC(2) results. 
These excitations have n-$\pi^*$ character and are blue-shifted due to the interactions with the environment. 
The magnitude of the reference shift falls in the 0.15-0.6~eV range, which makes the shift in these complexes  a suitable observable for discussing the effect of  the $\rho_B$-dependency of the FDET results. 
For this type of excitations, the combined effect of the approximation used for the FDET embedding potential and the use of the isolated environment density as $\rho_B(\vec{\mathrm{r}})$ results in the average error in the excitation energy of the magnitude of 0.04~eV~\cite{Zech2018,Ricardi2018}.

We start with the analysis of the results obtained without taking any experimental information from the molecular crystal, i.e., $\Delta\epsilon_{emb}[\rho_B^{\lambda_J=0}]$. 
$\Delta\epsilon_{emb}[\rho_B^{\lambda_J=0}]$ correspond to a "standard" FDET protocol in which the Hartree-Fock density of the isolated environment is used as $\rho_B(\vec{\mathrm{r}})$.  The deviation from the reference are small and their magnitude is consistent with the benchmark results published elsewhere.~\cite{Zech2018}. 
The effect of correlation on $\rho_B(\vec{\mathrm{r}})$ (see the shifts obtained with $\rho_B^{CCSD}(\vec{\mathrm{r}})$)  results in a slight reduction of the shifts in all cases.  

At $\lambda_J>0$, both the correlation- and the crystal-field effects are taken into account in $\rho_B(\vec{\mathrm{r}})$ leading to a further reduction of the shifts. The deviations of  FDET shifts from the reference increase (see the values of $\Delta\epsilon_{emb}[\rho_B^{\lambda_J=1}]$ in Figure~\ref{Figure_diagonal}).

As previously mentioned,we could not extend the restraint to values of $\lambda_J$ larger than 1, because the procedure became numerically unstable.
 ~\cite{Genoni2018}.
 Unfortunately, although the effect of correlation and polarisation by the crystal field are reflected in $\rho_B^{\lambda_J}(\vec{\mathrm{r}})$, they cannot be separated. 
 Moreover, the environment of GlyGly in the molecular crystal and in the clusters analysed in the present work are different. 
As a result, even if the reconstruction of the density of GlyGly from X-ray structure factors were exact, this would not guarantee that such density would yield the best FDET results for the clusters under investigation. 
The values 
  $\Delta\epsilon_{emb}[\rho_B^{\lambda_J}]$  at $\lambda_J=0$ and $\lambda_J=1$ represent, therefore, a good estimate of the maximum scatter (minimal and maximal bounds) of the FDET results due to the $\rho_B$-dependency of the FDET embedding potential.
 Within these bounds, the deviations from the reference do not exceed 0.1~eV (or 30\% in terms of the relative error).
This also points out the need for a thorough analysis of the disentangled effects of correlation and polarisation in $\Delta\epsilon[\rho_B^{\lambda_J}]$.

The subsequent part concerns the numerical stability of the FDET derived complexation induced shifts of the lowest excitation energy with respect to variations of $\rho_B(\vec{\mathrm{r}})$  correspondent to the change of the parameter $\lambda_J$ from 0 to 1.

\begin{figure}
\centering
\includegraphics[width=0.9\textwidth]{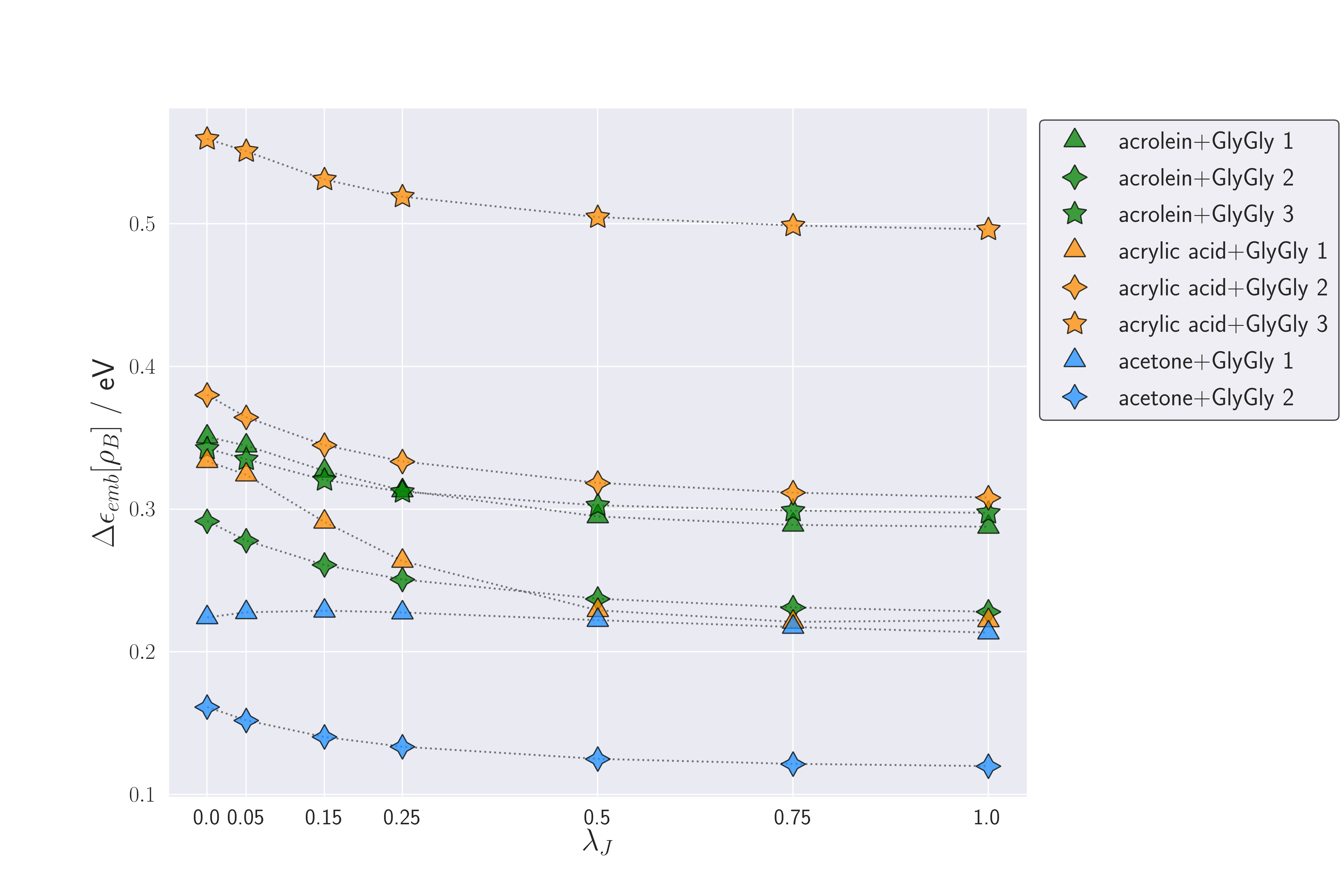}
\caption{Complexation induced shifts of the excitation energy  ($\Delta\epsilon_{emb}[\rho_B^{\lambda_J}]$) at  various values of $\lambda_J$ for 8 clusters.}
% acrolein+GlyGly 1, acrolein+GlyGly 2, acrolein+GlyGly 3 (green triangles, 4-point stars, and 5-point stars respectively), acrylic acid+GlyGly 1, acrylic acid+GlyGly 2, acrylic acid+GlyGly 3 (orange triangles, 4-point stars, and 5-point stars respectively), acetone+GlyGly 1, acetone+GlyGly 2 (light blue triangles and 4-point stars respectively).
\label{lambda_dep}
\end{figure}

Figure~\ref{lambda_dep} shows the dependence of the calculated shifts $\Delta\epsilon_{emb}[\rho_B^{\lambda_J}]$ on the parameter $\lambda_J$ for each  complex. The dependence of $\Delta\epsilon_{emb}[\rho_B^{\lambda_J}]$ on $\lambda_J$ is smooth and monotonic. 
Above $\lambda_J = 0.5$ till its maximal value used in this study $\lambda_J=1$, $\Delta\epsilon_{emb}[\rho_B^{\lambda_J}]$ 
remains almost constant (it changes by as little as about 0.01~eV).
\textcolor{black}{The magnitude of the solvatochromic shifts decreases for all but one system (acetone + GlyGly 1) when $\lambda_J$ increases. This can be ascribed to a decrease in dipole moment magnitude for the XRW densities when $\lambda_J$ increases. A similar trend appears for correlated methods, which, as it is known, tend to yield lower dipole moments.} 

\begin{figure}
\centering
\includegraphics[width=0.9\textwidth]{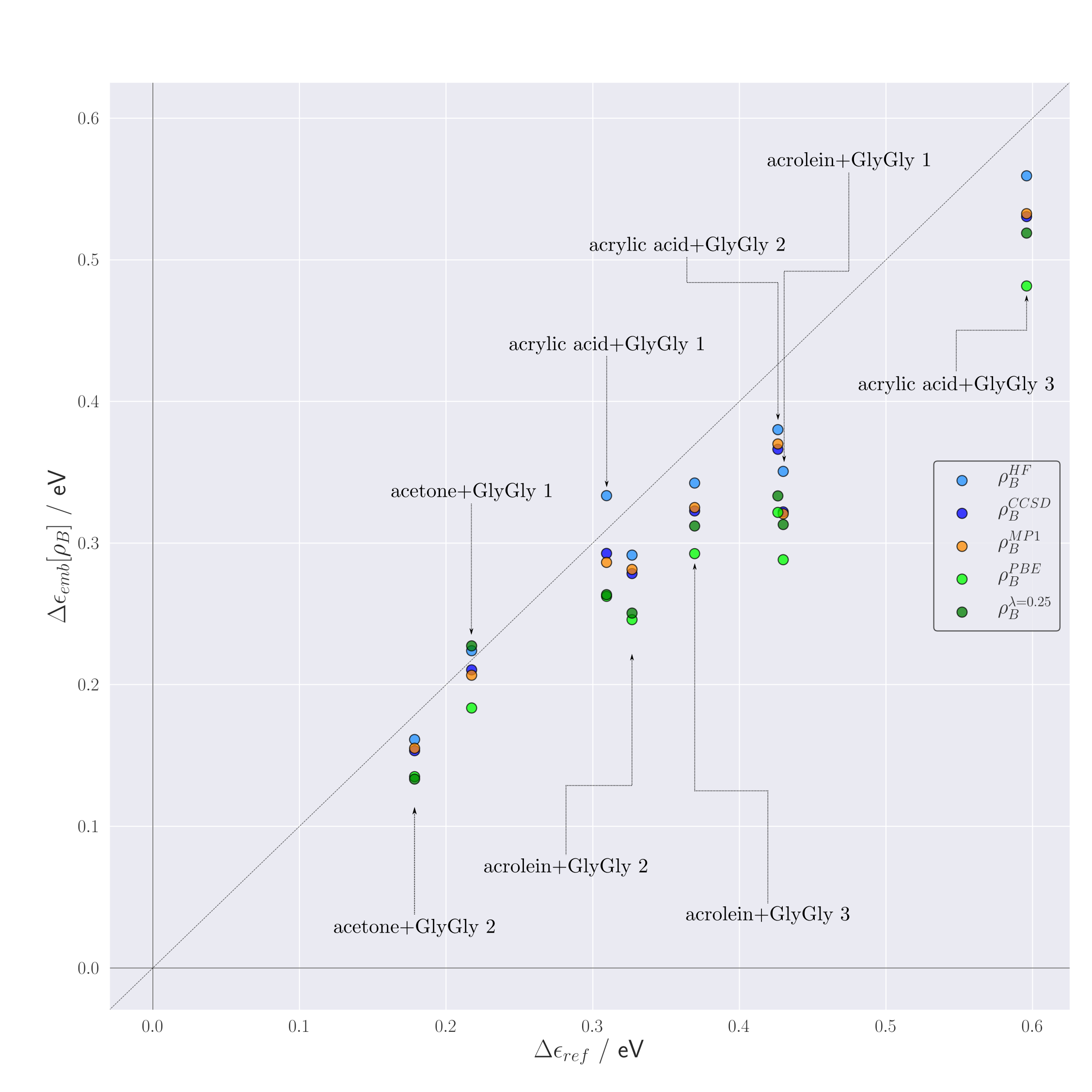}
\caption{FDET (embedded ADC(2)) derived complexation induced shifts of the  excitation energy ($\Delta\epsilon_{emb}$) obtained for eight intermolecular complexes with different choices for $\rho_B(\vec{\mathrm{r}})$.
%: $\rho_B^{MP}$ (green circles), $\rho_B^{HF}$ (blue circles), $\rho_B^{CCSD}$ (red circles), $\rho_B^{HF}$ (blue circles), $\rho_B^{PBE}$  (green diamonds), and  $\rho_B^{\lambda_J=0.25}$ (black full circles). 
Reference values ($\Delta\epsilon^{ref}$) are obtained from ADC(2) calculations for the whole complex.
}
\label{Figure_diagonal2}
\end{figure}

Turning back to practical applications, we notice that large scale simulations usually apply  the monomer expansion
 for both 
 $\rho_A(\vec{\mathrm{r}})$ and $\rho_B(\vec{\mathrm{r}})$ and using isolated environment density as 
 $\rho_B(\vec{\mathrm{r}})$  already assures good accuracy of the FDET derived environment induced shifts. 
In such simulations, the modeller has a wide range of available methods 
 to generate $\rho_B(\vec{\mathrm{r}})$ (see the Introduction).
 The data collected in  Figure \ref{Figure_diagonal2},  shows how the FDET results depend on the method used to generate  $\rho_B(\vec{\mathrm{r}})$ including: Hartree-Fock, MP1, CCSD, and KS-DFT(PBE). 
For reference purposes, the values of $\Delta\epsilon_{emb}$
obtained from X-ray diffraction data at $\lambda_J=0.25$ are also given. 

The results collected in Figure~\ref{Figure_diagonal2} indicate clearly that X-ray derived molecular densities are suitable  for generating $\rho_B(\vec{\mathrm{r}})$ for FDET calculations following the conventional protocol (LinearizedFDET, 
monomer expansion of $\rho_A(\vec{\mathrm{r}})$, monomer expansion for $\rho_B(\vec{\mathrm{r}})$, lack of explicit treatment of $\rho_B(\vec{\mathrm{r}})$ polarisation by the chromophore). The deviations from the reference are, however, larger than if Hartree-Fock or correlated isolated GlyGly densities are used for this purpose. 
This does not bear direct relevance to the quality of these densities. The overall error of the FDET derived excitation energy results  from the balance of the errors in two FDET embedding potentials evaluated at two different pairs of densities  $v_{emb}^{}[\rho_{A}^{ES},\rho_{B}^{},v_{B}^{}](\vec{\mathrm{r}})$ and $v_{emb}^{}[\rho_{A}^{GS},\rho_{B}^{},v_{B}^{}](\vec{\mathrm{r}})$, where $\rho_{A}^{GS}$ and  $\rho_{A}^{ES}$ denote ground and excited state, respectively, in which the non-electrostatic contributions are approximate.

It is worthwhile to note that the use of X-ray derived densities as  $\rho_B(\vec{\mathrm{r}})$  leads to smaller errors than if the Kohn-Sham PBE calculations are used for this purpose. 

\section{Conclusions}

The recent developments in techniques to reconstruct the electron density from X-ray diffraction data made it possible not only to determine the maxima of the electron density (coinciding with the position of nuclei) but also to reveal its more detailed features~\cite{Genoni2018}. 
For a molecular crystal, the reconstruction yields a localised density of a single molecule but taking into account its chemical environment.
In FDET based simulations, the molecular densities are used as an input quantity providing the complete quantum-mechanical descriptor of the environment of the embedded species.
In the present work, we explored the possibility to use molecular density of the glycylglycine derived from X-ray diffraction data collected for the molecular crystal in FDET calculations of complexation induced shift of the excitation energy in eight intermolecular complexes, each consisting of an organic chromophore hydrogen-bonded to one glycylglycine molecule.  The usability of such densities for this purpose was not evident before the present study was made. Several factors could, in principle, invalidate such practical applications of X-ray reconstructed densities.  First of all, glycylglycine in the crystal and in the complexes analysed in the present work have different environments. This  might result in different polarisation of such localised molecular, and as a consequence, contribute to errors in the FDET results.   Other group of factors relate rather to the reconstruction procedure. It cannot be made perfect due to a) errors in the experimental measurements, b) the very basic assumption according to which the average of a dynamic quantity (electron density) is represented using as an intermediate object, namely a static single-determinant wave-function, c) incompleteness of the experimental data, d) errors in the phasing procedures, e) crystal defects.
It could be expected, therefore, that reconstructed densities deviate indeed not significantly from the constraint-free one and, in turn,
yield similar excitation energies  if used as $\rho_B(\vec{\mathrm{r}})$. 
The primary objective of this work was the verification of this expectation. 
The obtained results demonstrate, indeed, that X-ray reconstructed densities are suitable to be used as $\rho_B(\vec{\mathrm{r}})$ in FDET on par with possible alternative techniques. 

Despite the fact that the X-ray restrained wavefunction procedure does not yield a unique solution but rather a range of densities parametrised by $\lambda_J$, the scatter of the excitation energies obtained using the whole range of this parameter is rather narrow. \textcolor{black}{For all but one complex (acetone~+~GlyGly~1, characterised by a remarkably short hydrogen bond of only 1.3~$\AA$) the excitation energies vary within approximately 0.05~eV depending on the details of the reconstruction procedure.
This scatter in calculated shifts is small compared to the range of variation of solvatochromic shifts~\cite{Reichardt1994,Improta2016} making FDET simulations using X-ray derived molecular densities an attractive tool for making quantitative predictions and to interpret experimental results.}
Further reduction of this scatter is probably possible through disentangling the effects of crystal-field polarisation and correlation effect on the density of a molecule in a molecular crystal. We intend to deal with this issue in our subsequent work.
Also here the experiment-derived $\rho_B(\vec{\mathrm{r}})$ might prove more useful than alternative techniques. This is the case when the molecules associated with $\rho_B(\vec{\mathrm{r}})$ have similar neighbours in the cluster to be investigated and in the molecular crystal used to generate  $\rho_B(\vec{\mathrm{r}})$.

\textcolor{black}{
For the first time, we adopted an experimental density for the environment and tested this on spectral shifts for valence excitations. At present, a density calculated via an X-ray restrained procedure is the only possibility for this approach.
The numerical examples in this work in which we applied the proposed procedure concern microsolvated clusters. 
For finite systems, many alternatives to generate $\rho_B(\vec{\mathrm{r}})$ involving similar or even lower computational cost are possible  (Hartree-Fock or Kohn-Sham densities of isolated environment including or not their optimization or pre-polarization).
This numerical validation is the first stage in our long-standing interests and plans aiming at modelling the electronic structure of species in the condensed phase such as neat or doped molecular crystals~\cite{Meirzadeh2018}. 
% In the present work, the applicability of experiment-derived $\rho_B(\vec{\mathrm{r}})$  in FDET was tested on spectral shifts for valence excitations.
We plan to apply the same strategy to generate the FDET embedding potential using experiment-derived $\rho_B(\vec{r)}$ for modeling other spectroscopic properties (core excitations, NMR shifts, two-photon absorption, hyperpolarizabilities, etc.) that are evaluated from embedded wavefunctions.
In infinite systems, the generation of $\rho_B(\vec{\mathrm{r}})$ from first principles might face serious difficulties if made in a similar way as for the studies of clusters,  for the reasons listed below.
Firstly, a density obtained using  a straightforward application of the simplest protocol (generation of $\rho_B(\vec{\mathrm{r}})$  in an artificial system with a void in place of the part described by means of $\Psi_A$) might be  unphysical or even impossible to obtain due to the convergence problems or the need of a much larger supercell.
Secondly, taking into account electronic correlation on electron density  in periodic systems is generally limited to Kohn-Sham type of methods.
A final point is worth discussing. By using a wavefunction restrained to fit the electron density of a molecule
in a crystal, the environment density used in the approach we proposed includes implicitly not only the effects of intermolecular
interactions but also long range electrostatic effects of the crystalline matter.}
%The proposed protocol not only makes the generation of $\rho_B(\vec{\mathrm{r}})$ a rather low-cost task, but assures also that the effect of the crystal field on the density of each molecule composing the total $\rho_B(\vec{\mathrm{r}})$  is taken into account.

\subsection{Synopsis}
We demonstrate -for the first-time- the use of experiment-derived molecular electron densities as $\rho_B(\vec{\mathrm{r}})$ in Frozen Density Embedding Theory based calculations of environment-induced shifts of electronic excitations for chromophores in clusters. $\rho_B(\vec{\mathrm{r}})$ was derived from X-ray restrained molecular wavefunctions of glycylglycine in molecular crystals to obtain environment densities for modelling the clusters.

\subsection{Acknowledgement}
 This research was supported by grant from the Swiss National Science Foundation (Grant No. 200020-172532).
 
\bibliography{group_bibliography}
\bibliographystyle{ieeetr}

\end{document}